\documentclass{elsart}
\usepackage{amssymb}
\usepackage{graphicx}

\usepackage[ps2pdf,colorlinks=true]{hyperref}
 \def\ket{\!>\,}

\begin{document}
\begin{frontmatter}

\title{Investigation of doublet-bands in 
$^{124,126,130,132}$Cs odd-odd nuclei using triaxial projected shell model approach}
\author{G.H. Bhat$^{1}$, R.N. Ali$^{1}$, J.A. Sheikh$^{1,2}$ and R. Palit$^{3}$ }
\address{
$^1$Department of Physics, University of Kashmir, Srinagar,
190 006, India \\
$^2$Department of Physics and Astronomy, University of
Tennessee, Knoxville, TN 37996, USA\\
$^3$Department of Nuclear and Atomic Physics, Tata Institute of Fundamental Research, Colaba, Mumbai, 400 005,India}
\begin{abstract}
Doublet bands observed in $^{124,126,130,132}$Cs isotopes are studied 
using the recently developed multi-quasiparticle microscopic
triaxial projected shell model (TPSM) approach. It is shown that TPSM results for energies
and transition probabilities are in good agreement with known energies and the recently 
measured extensive 
data on transition probabilities for the bands in $^{126}$Cs. In particular, it is demonstrated that
characteristics transition probabilities expected for the doublet bands to originate from the
chiral symmetry breaking are well reproduced in the present work. The calculated energies for 
$^{124,130,132}$Cs are also shown to be in reasonable agreement with the available experimental data.
Furthermore, a complete set of the calculated transition probabilities is provided for the doublet
bands in $^{124,130,132}$Cs isotopes. 

\end{abstract}
\end{frontmatter}


\section{Introduction}
The observation of chiral rotation in atomic nuclei has attracted a considerable attention
in recent years. It has been demonstrated that band structures of triaxial odd-odd nuclei 
in $A \sim 100$ and $\sim 130$ regions show characteristic properties of chiral symmetry breaking
in the intrinsic frame \cite{VI00,SK02,PO04,CV04}. 
In the $A \sim 130$ region, odd-proton (odd-neutron)
occupies lower-half (upper-half) of the intruder sub-shell $1h_{11/2}$ with their angular-momentum
vectors aligned toward short (long) principal-axis of the triaxial density distribution. The 
core angular-momentum vector, on the other hand, is directed along the intermediate axis
as it has the largest irrotational moment of inertia. The three orthogonal angular-momentum
vectors of odd-proton, odd-neutron and the core give rise to chiral geometry \cite{SF97}.
The manifestation
of the chiral geometry in the laboratory frame is the appearance of $\Delta I =1$ doublet 
band structures with similar $\gamma$-ray energies \cite{KS01,TK03}. Although, near degenerate doublet
bands have been observed in many odd-odd nuclei as well as in odd-mass nuclei, however, 
it is now well recognised that several of these bands don't originate from the chiral
symmetry breaking  \cite{KK01,AA01,SU03,VD00,PRM,DT06}. The stringent selection rules imposed by chiral symmetry on the wavefunction
in the form of characteristic transition probabilities are not obeyed by many nuclei, although,
they depict near degenerate doublet bands\cite{CM06,DT07}. The measurement of transition probabilities, therefore,
plays a very critical role in the establishment of chiral geometry for a given system. 
Previously, $^{128}$Cs was the best known example where most of the selection rules on $\gamma-$transitions governed by spontaneous chiral symmetry breaking were approximately satisfied 
\cite{EG06}. For this
nucleus, the energy separation between the doublet bands is approximately $200~keV$ and the energy 
staggering is very small. ${B(E2)}(I \rightarrow I-2)$ transitions were measured up to spin, 
$I=20$ for the yrast and up to $I=17$ for the side band. These ${B(E2)}$ values for the two bands
are quite similar, which confirms that doublet bands originate from the same intrinsic
configuration. ${B(M1)}(I \rightarrow I-1)$ transitions for 
yrast$\rightarrow$yrast, side$\rightarrow$side and side$\rightarrow$yrast were measured 
up to spins of $I=18, 17$ and 16, respectively. It is evident from the data that ${B(M1)}$ 
transitions for side$\rightarrow$yrast have staggering phase opposite to that of 
yrast$\rightarrow$yrast and side$\rightarrow$side phase, which is a result of the
selection rule originating from the chiral symmetry. These observed properties
of the chiral bands in $^{128}$Cs have been studied using the triaxial particle-rotor model \cite{EG06}
and also were investigated recently using the triaxial projected Shell model (TPSM)
approach \cite{JG12}. It was demonstrated that TPSM gives a slightly better description of the observed
data as compared to the particle-rotor model approach. In particular, it was shown that 
TPSM correctly reproduces the observed trend in the measured ${B(E2)}$ transitions as a 
function of spin.

The purpose of the present work is to perform a systematic investigation of the observed 
doublet band structures in odd-odd Cs-isotopes using the TPSM approach. Recently, 
using the doppler shift attenuation (DSA) method, 13 lifetimes and 26 corresponding 
absolute transition
probabilities were obtained for $^{126}$Cs \cite{EG12}. This set of data on transition 
probabilities 
is more exhaustive than that of $^{128}$Cs and it is possible to test the spontaneous
breaking of chiral symmetry more completely. For other Cs-isotopes, studied in the present
work, limited data for the doublet-bands is available for the transition energies only
and we provide a complete set of transition probabilities for these nuclei
using the TPSM approach to be compared with the future experimental measurements.

The present manuscript is organized as follows. In section II, we shall present a few
elements of the TPSM approach that are relevant in the
discussion of the results. The details of the TPSM approach have already been presented in our 
earlier publications, for instance, in  ref. \cite{JG12}. The obtained results are displayed
and discussed in section III
and finally summary and conclusions are provided in section IV.

\section{Triaxial projected shell model approach}
For odd-odd nuclei, the TPSM basis space consists of one-neutron plus one-proton quasiparticle
configurations \cite{RJ03,RJ01} :
\begin{equation}
\{ | \phi_\kappa \ket = {a_\nu}^\dagger {a_\pi}^\dagger | 0 \ket \}  .
\label{intrinsic}
\end{equation}
As almost all the doublet bands observed in odd-odd nuclei are built on one-neutron plus one-proton
configurations, the above chosen basis is adequate to discuss the doublet band structures. In
Eq.~(\ref{intrinsic}), $| 0 \ket $ is the quasiparticle vacuum state and is obtained through
diagonalization of the deformed Nilsson Hamiltonian with a subsequent
BCS calculations. The number of basis configurations in Eq.~(\ref{intrinsic}) 
depend on the number of
levels near the respective Fermi levels of protons and neutrons. 

It should  be noted that the states $ | \phi_\kappa \ket $ constructed from the diagonalization
of the deformed Nilsson 
Hamiltonian do not conserve rotational symmetry. In order to  
restore this symmetry, three-dimensional angular-momentum projection technique 
is applied \cite{ringshuck}. 
From each intrinsic state, $\kappa$, in (\ref{intrinsic}) a band is generated through 
projection technique. The interaction 
between different bands for a given spin is taken into account by diagonalizing
the shell model Hamiltonian in the projected basis. 
The Hamiltonian employed in the present work is given by
\begin{equation}
\hat{H} = \hat{H_0} - \frac{1}{2} \chi \sum_\mu  \hat{Q}_\mu^\dagger 
\hat{Q}_\mu - G_M \hat{P}^\dagger \hat{P} 
- G_Q \sum_\mu \hat{P}_\mu^\dagger \hat{P}_\mu,
\label{Hamilt}
\end{equation}
and the corresponding triaxial Nilsson Hamiltonian is given by
\begin{equation}
\hat H_N = \hat H_0 - {2 \over 3}\hbar\omega\left\{\epsilon\hat Q_0
+\epsilon'{{\hat Q_{+2}+\hat Q_{-2}}\over\sqrt{2}}\right\} ,
\label{nilsson}
\end{equation}
\begin{table}
\begin{center}
\caption{Axial deformation parameter ($\epsilon$) and triaxial
deformation parameter ($\epsilon'$) employed in the calculation for
$^{124,126,130,132}$Cs-isotopes. }
\begin{tabular}{c|cccc}
\hline            & $^{124}$Cs&$^{126}$Cs   & $^{130}$Cs  & $^{132}$Cs \\
\hline $\epsilon$ & 0.256    & 0.260      & 0.160      & 0.170      \\
       $\epsilon'$& 0.170    & 0.150      & 0.145      & 0.150    \\\hline
\end{tabular}
\end{center}
\end{table}

\begin{figure}[htb]
 \centerline{\includegraphics[trim=0cm 0cm 0cm
0cm,width=0.55\textwidth,clip]{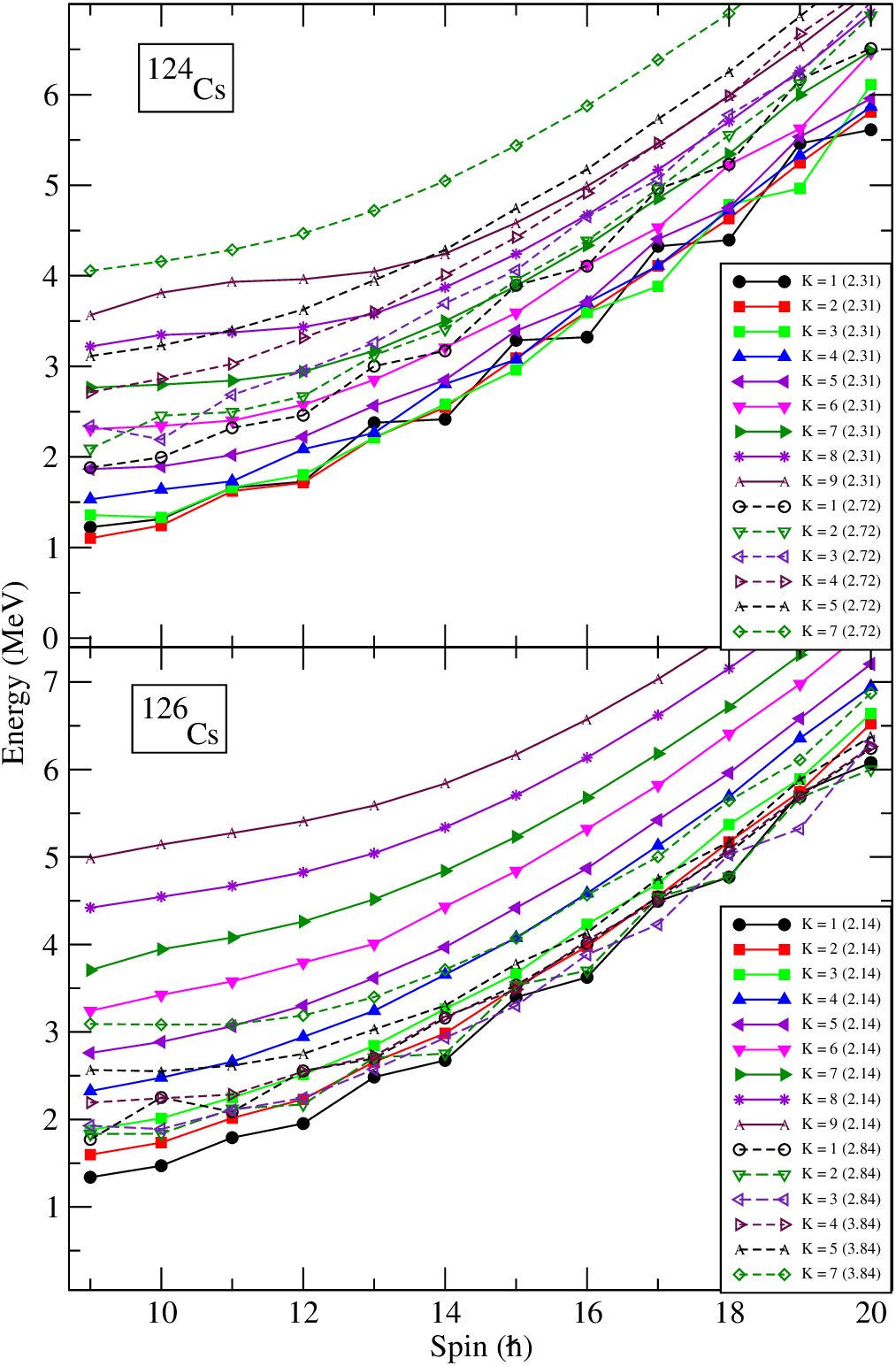}}
\caption{(Color online)  Angular-momentum projected bands obtained
 for different intrinsic K-configuration, given in the legend box, for $^{124,126}$Cs nuclei.
The energies of the triaxial quasiparticle states are given in the parenthesis.}
\label{fig1}
\end{figure}
\begin{figure}[htb]
 \centerline{\includegraphics[trim=0cm 0cm 0cm
0cm,width=0.55\textwidth,clip]{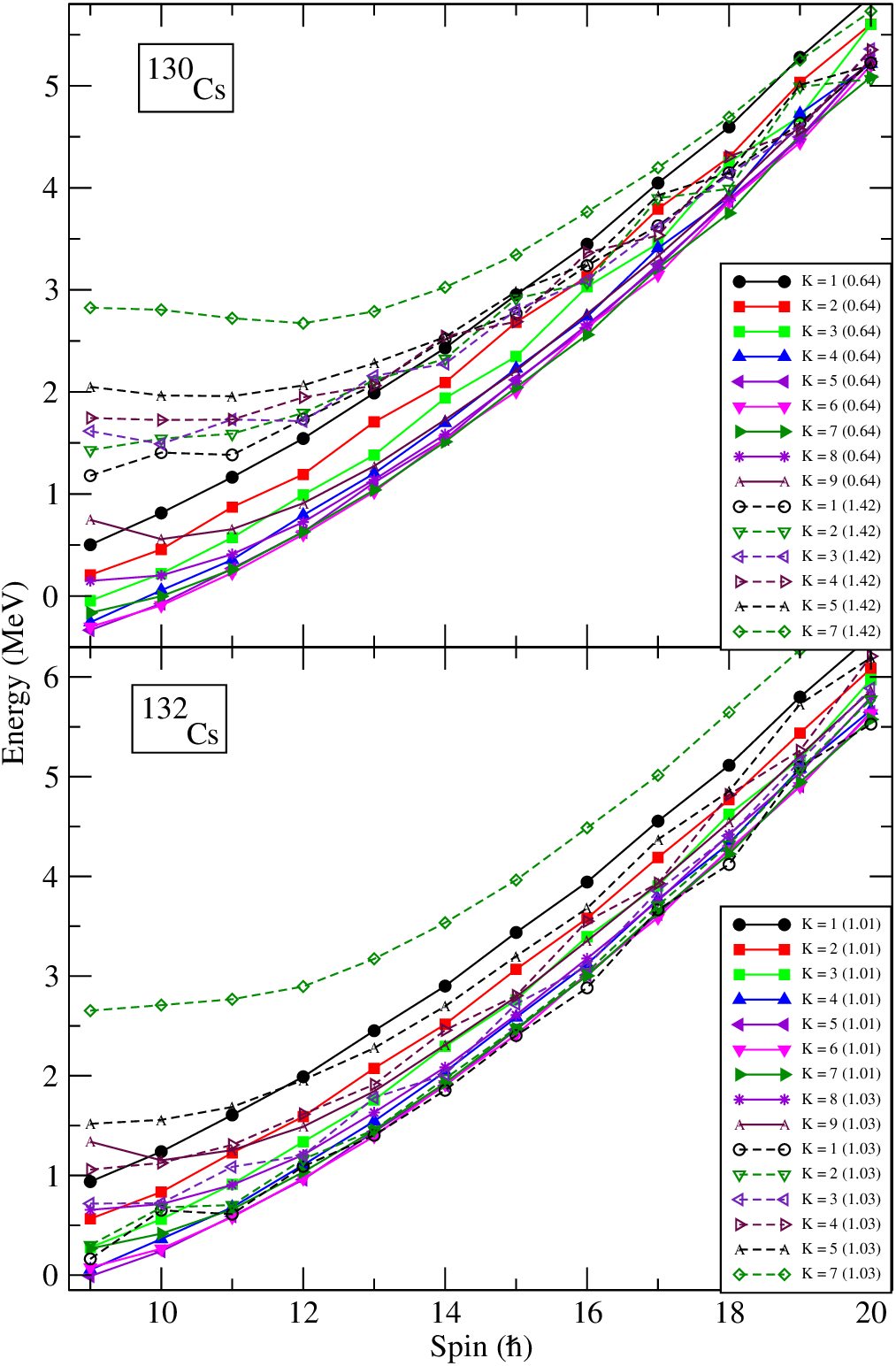}}
\caption{(Color online)  The angular-momentum projected bands obtained
 for different intrinsic K-configuration, given in legend box, for $^{130,132}$Cs nuclei.
The energies of the triaxial quasiparticle states are given in the parenthesis.}
\label{fig1}
\end{figure}
\begin{figure}[htb]
 \centerline{\includegraphics[trim=0cm 0cm 0cm
0cm,width=0.60\textwidth,clip]{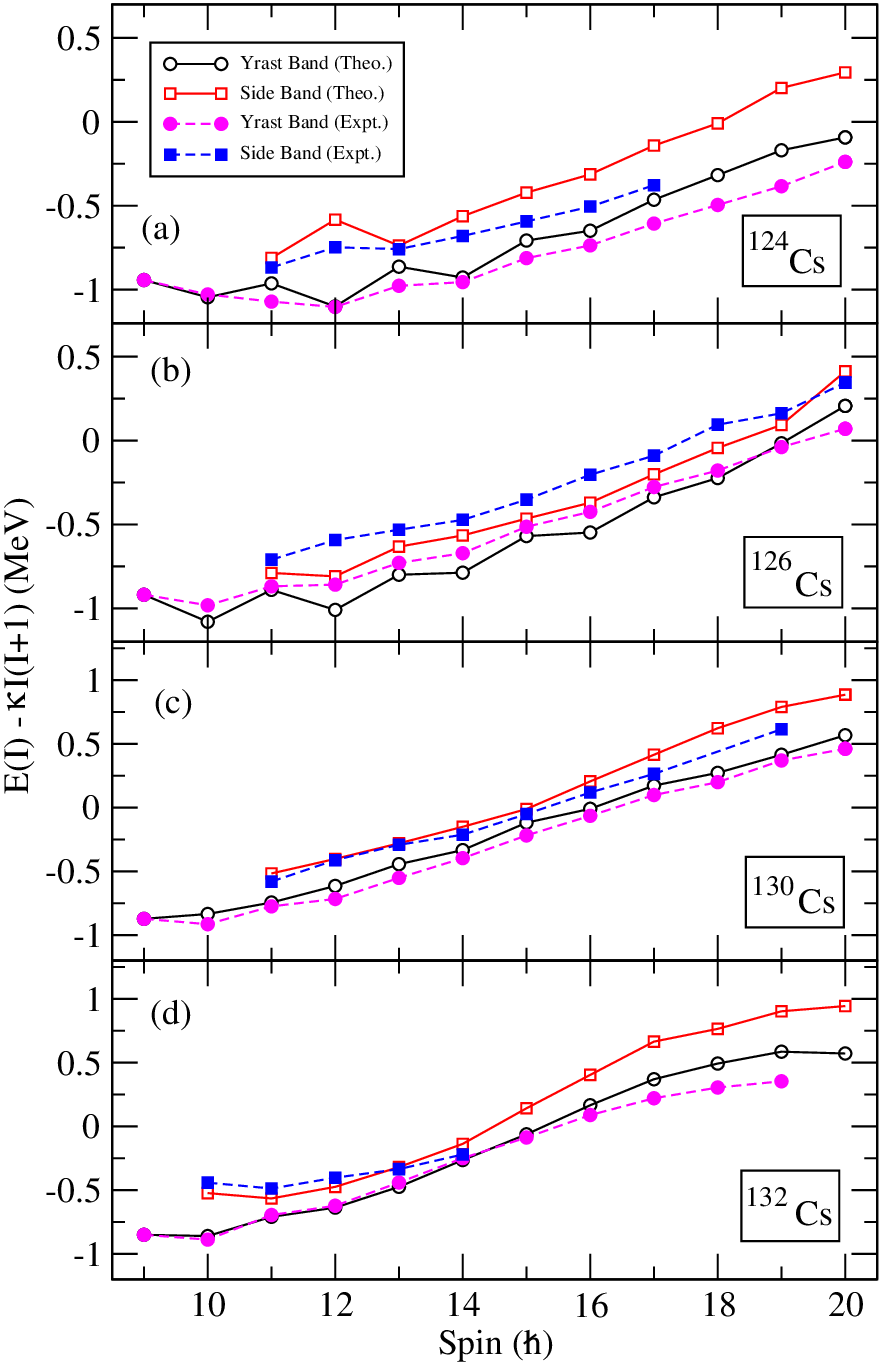}}
\caption{(Color online) Comparison of the TPSM energies, after configuration 
mixing with the available experimental data for the yrast and side bands of the
studied $^{124,126,130,132}$Cs nuclei. The value of $\kappa$, shown in y-axis, is defined as 
$\kappa=32.32 A^{-5/3}$. Data have been taken from Refs. 
\cite{EG12,AG01,GR03,Simons1}  }
\label{fig1}
\end{figure}
\begin{figure}[htb]
 \centerline{\includegraphics[trim=0cm 0cm 0cm
0cm,width=0.60\textwidth,clip]{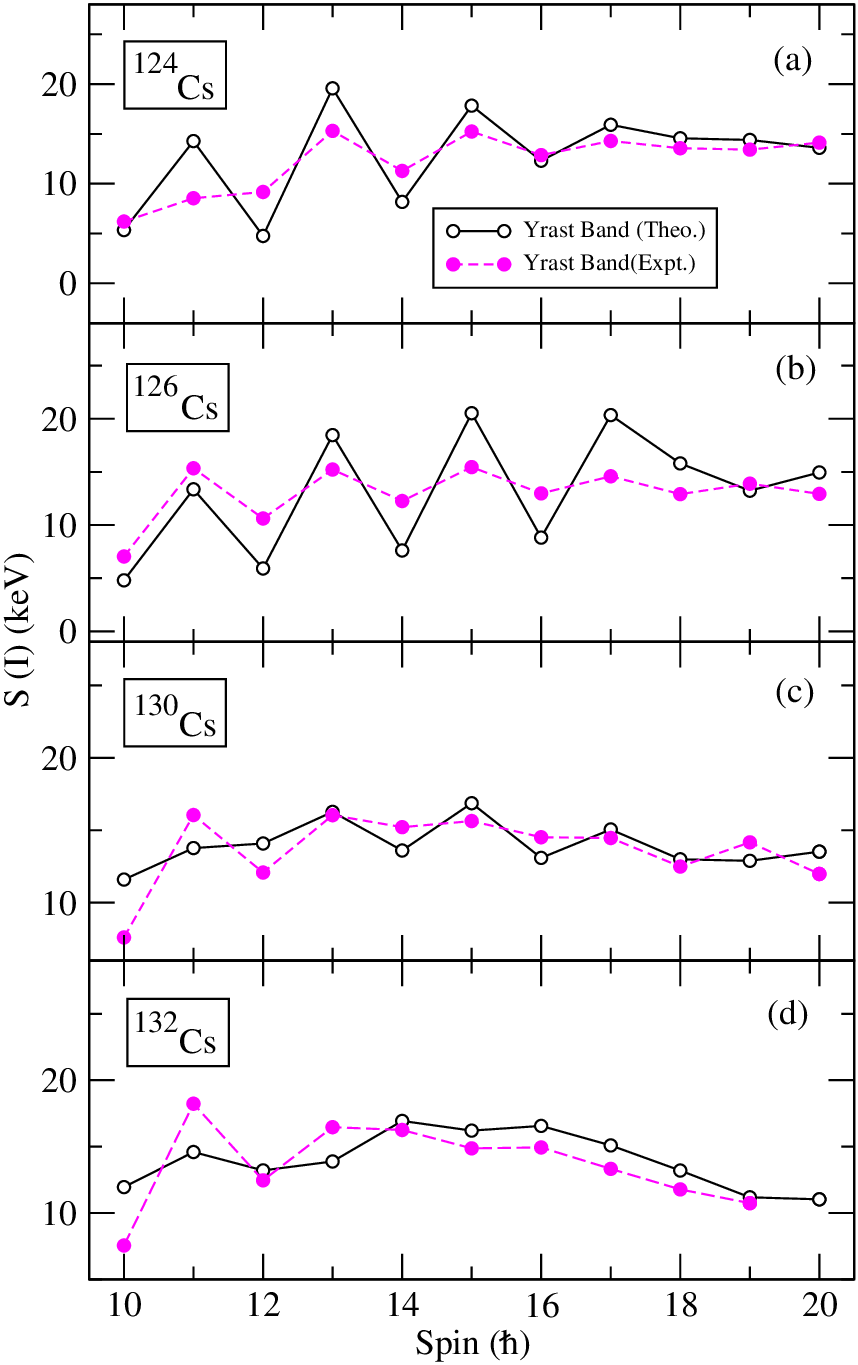}}
\caption{(Color online) 
 Calculated staggering 
parameter $S(I)=\left[E(I)-E(I-1)\right]/2I $ is plotted along with the
 measured values in  panels.  }
\label{fig1}
\end{figure}
\begin{figure}[htb]
 \centerline{\includegraphics[trim=0cm 0cm 0cm
0cm,width=0.60\textwidth,clip]{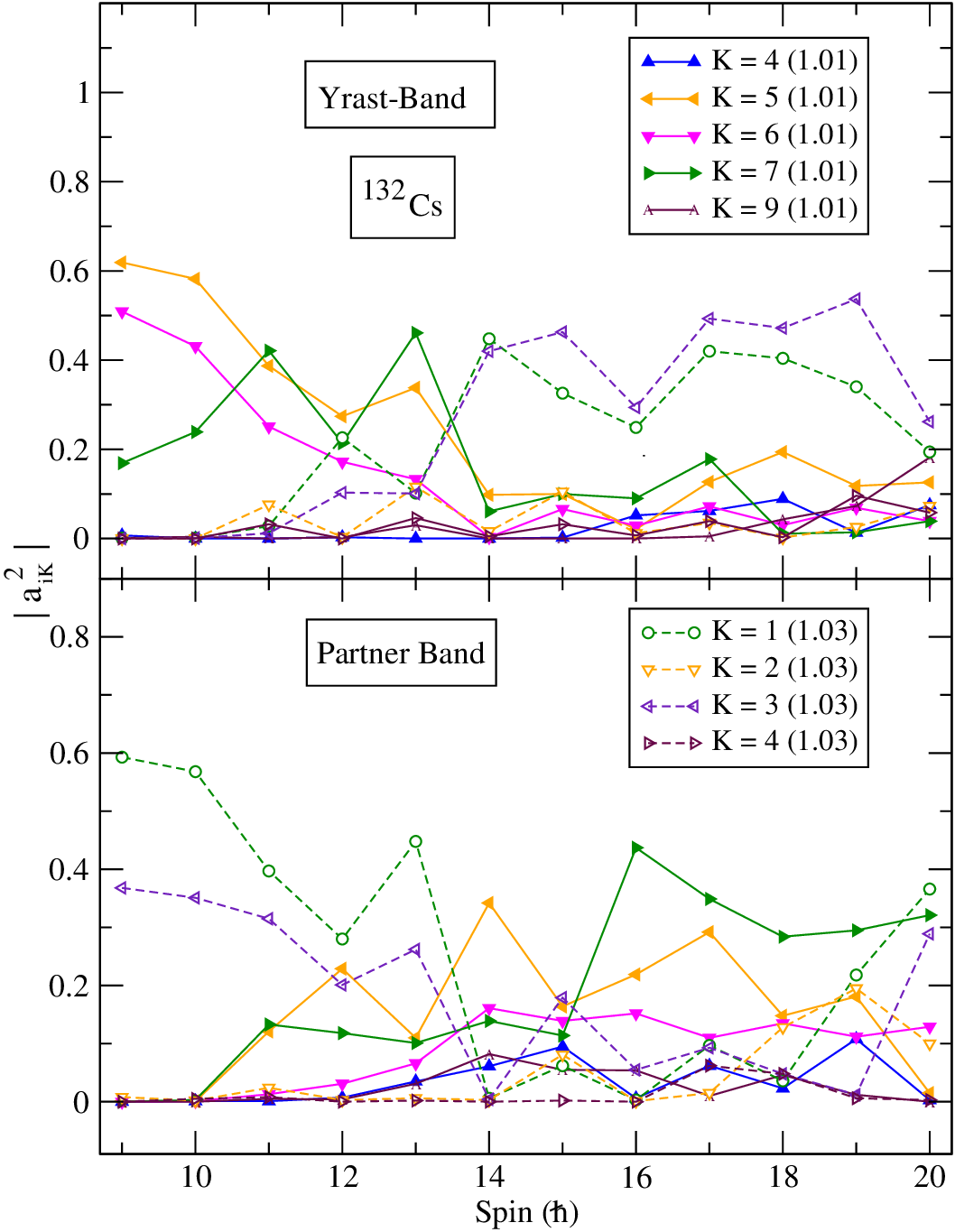}}
\caption{(Color online) Probability of various projected K-configurations in the wavefunctions
of the yrast and the first excited  bands for $^{132}$Cs }
\label{fig1}
\end{figure}

\begin{figure}[htb]
 \centerline{\includegraphics[trim=0cm 0cm 0cm
0cm,width=0.60\textwidth,clip]{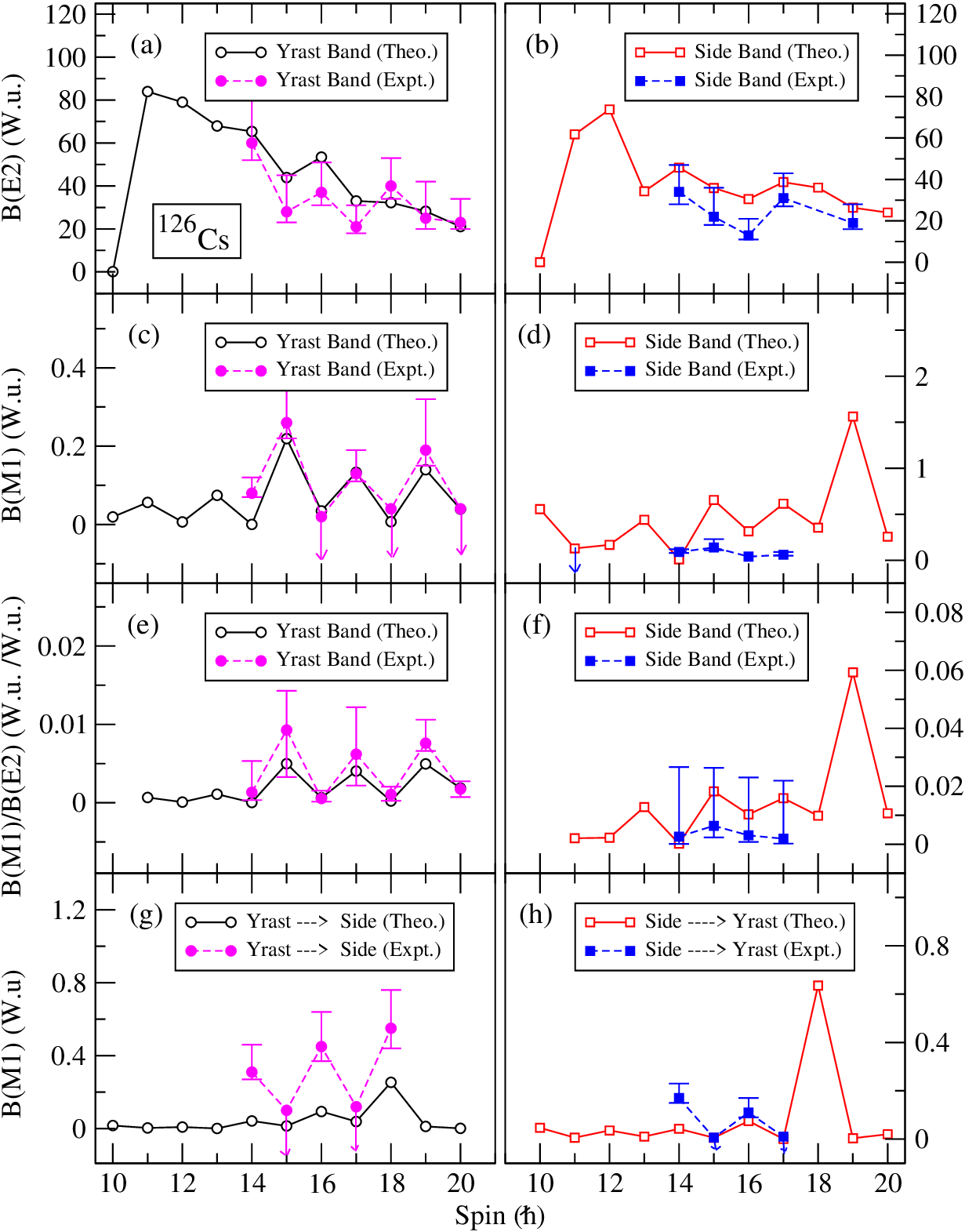}}
\caption{(Color online) Comparison of the experimental B(E2), B(M1) 
transition strengths and their ratios for $^{126}$Cs \cite {EG12} with the 
calculated values for yrast and side bands are depicted in panels
(a) to (f). The calculated inter-band transitions are shown in panels (g)
 and (h), and these panels also depicts the measured values for a 
few transitions. }
\label{fig1}
\end{figure}
\begin{figure}[htb]
 \centerline{\includegraphics[trim=0cm 0cm 0cm
0cm,width=0.60\textwidth,clip]{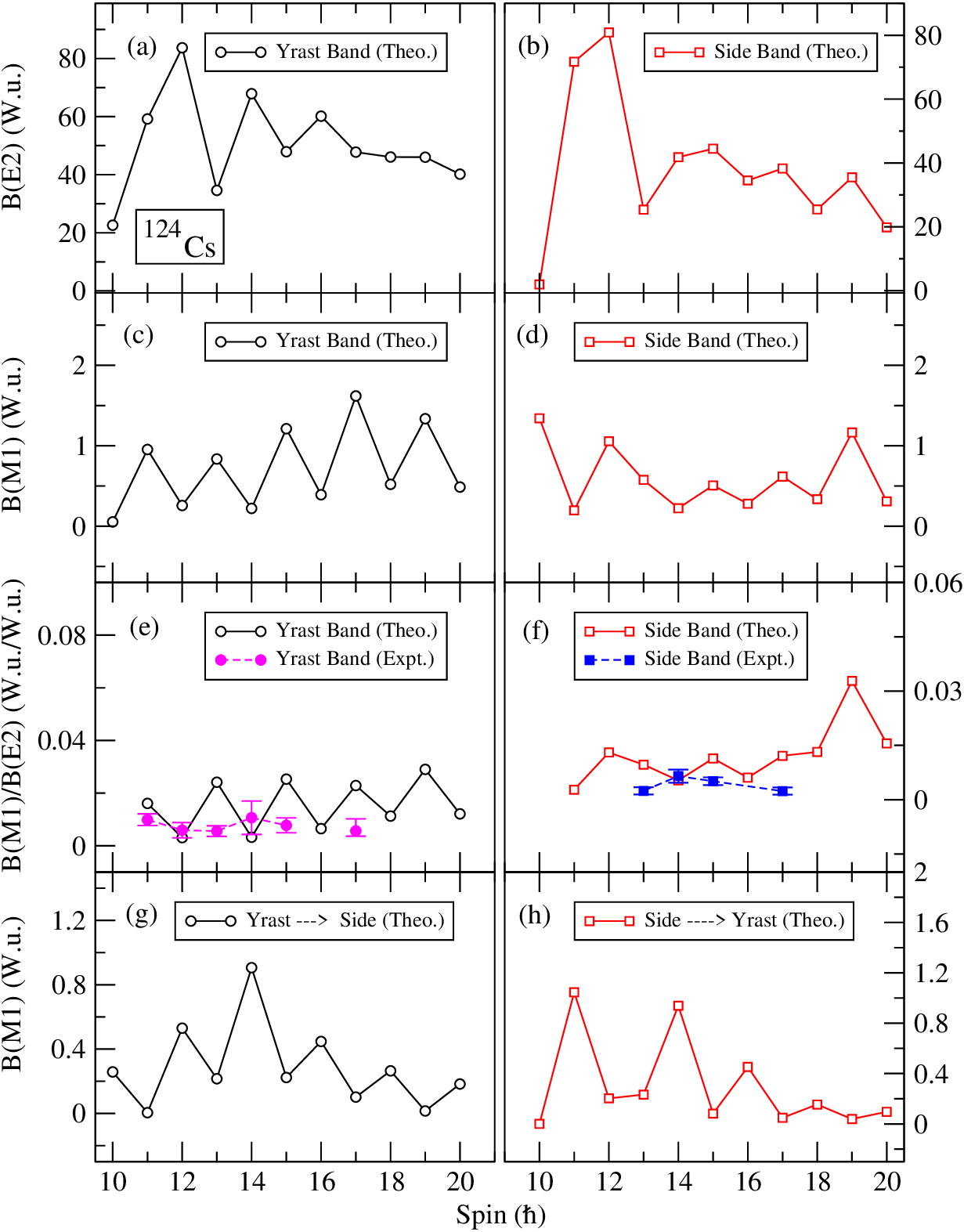}}
\caption{(Color online) The calculated B(E2), B(M1)
transition strengths and their ratios for $^{124}$Cs are depicted in panels
(a) to (f). The measured ratios \cite{AG01} are also shown in the panels (e) and (f)
The calculated inter-band transitions are shown in panels (g)
 and (h).}
\label{fig1}
\end{figure}
\begin{figure}[htb]
 \centerline{\includegraphics[trim=0cm 0cm 0cm
0cm,width=0.60\textwidth,clip]{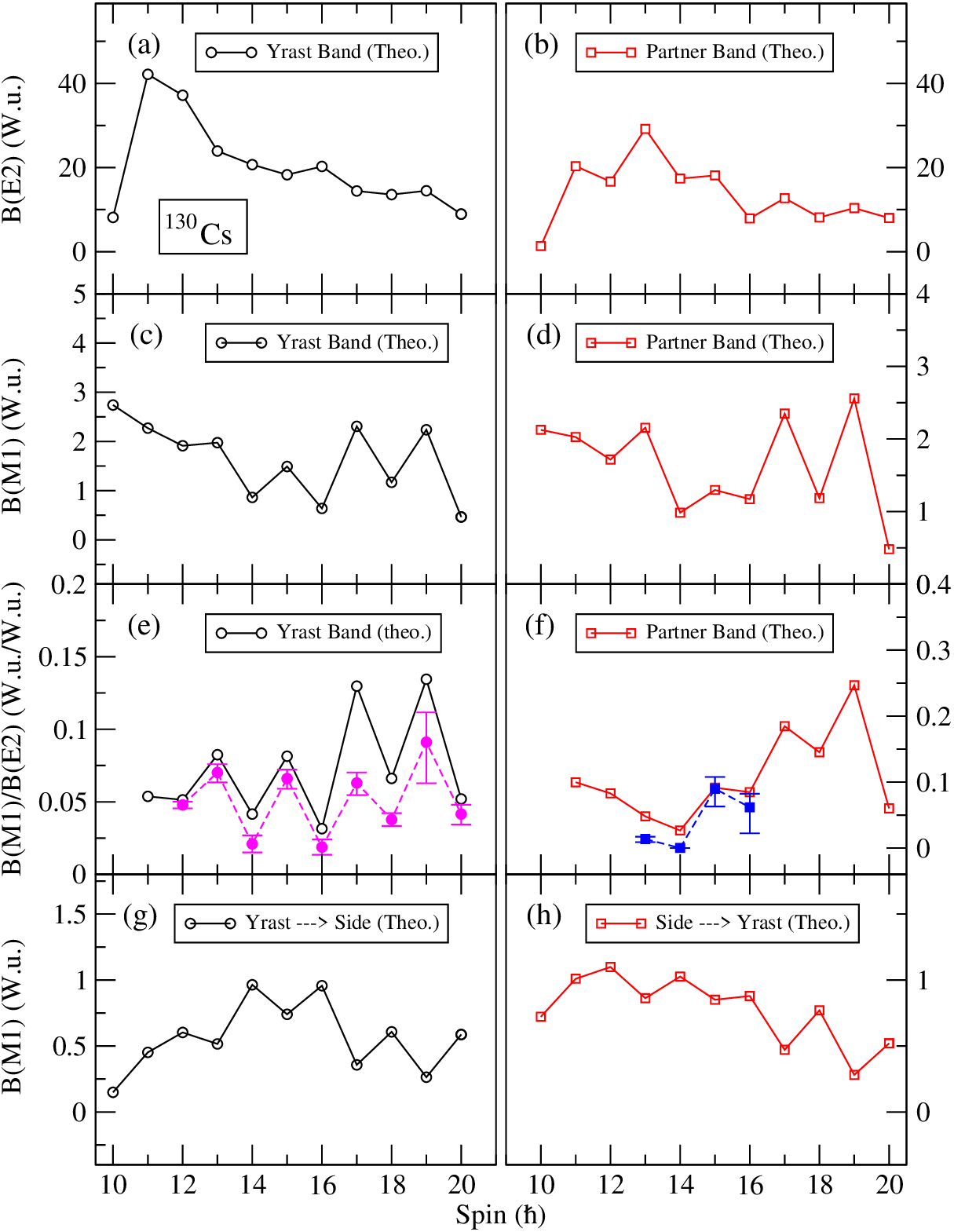}}
\caption{(Color online) The calculated B(E2), B(M1) 
transition strengths and their ratios for $^{130}$Cs are depicted in panels
(a) to (f). The measured ratios \cite{Simons1} are also shown in the panels (e) and (f)
The calculated inter-band transitions are shown in panels (g)
 and (h).}
\label{fig1}
\end{figure}
\begin{figure}[htb]
 \centerline{\includegraphics[trim=0cm 0cm 0cm
0cm,width=0.60\textwidth,clip]{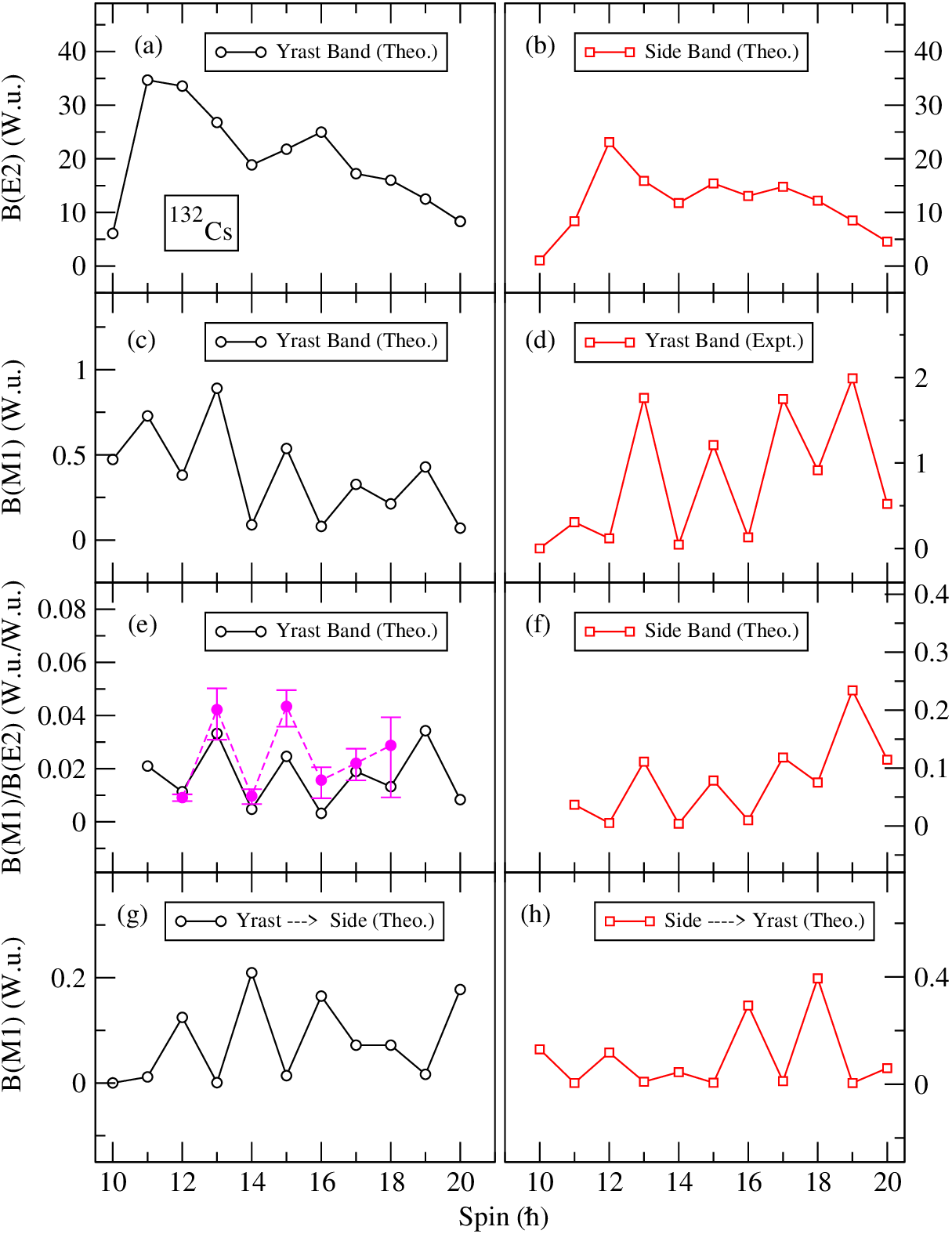}}
\caption{(Color online)The calculated B(E2), B(M1)
transition strengths and their ratios for $^{132}$Cs are depicted in panels
(a) to (f). The measured ratios \cite{GR03} are also shown in the panels (e) and (f)
The calculated inter-band transitions are shown in panels (g)
 and (h). }
\label{fig1}
\end{figure}
where $\hat{H_0}$ is the spherical single-particle shell model Hamiltonian,
which contains the spin-orbit force \cite{Ni69}. The second, third 
and fourth terms in Eq.~(\ref{Hamilt}) represent quadrupole-quadrupole, 
monopole-pairing, and quadrupole-pairing interactions, respectively. 
The axial and triaxial terms of the Nilsson potential in
Eq. \ref{nilsson} contain the parameters $\epsilon$ and $\epsilon'$, 
respectively, which are approximately related
to the $\gamma$-deformation parameter by 
$\gamma$ = tan$^{-1}(\frac{\epsilon'}{\epsilon})$. The strength of the 
quadrupole-quadrupole force $ \chi $ is 
determined in such a way that the employed quadrupole deformation $ \epsilon $
is same as obtained by the Hartree-Fock-Bogoliubov (HFB) procedure. 
The monopole-pairing force constants $G_M$ used in the calculations are
\begin{equation}
G_M ^\nu = \lbrack 20.12 - 13.13 \frac{N-Z}{A} \rbrack A^{-1}, ~~~ 
G_M ^\pi = 20.12 A^{-1} . 
\end{equation}
The quadrupole pairing strength $G_Q$ is assumed to be proportional to
the monopole strength,
$G_Q = 0.16 G_M$. All these interaction strengths are the same as those
used in our earlier studies \cite{JG12,RJ03,RJ01}. 

\section{Results and discussions}
The investigation of the properties of a triaxial system using the TPSM approach 
proceeds in several stages. In the
first stage, the triaxial intrinsic states are constructed by solving the three-dimensional
Nilsson potential with input axial and non-axial deformation
parameters. The deformations employed for the Cs-isotopes
investigated in the present work are listed in Table I. The 
axial deformation values have been adopted from the earlier studies \cite{KK01,AG01,SY10,JM06,SY07,GR03} 
and the chosen value of non-axial deformation approximately corresponds to $\gamma \sim 30^o$. 

In the second stage, 
three-dimensional angular-momentum projection technique is used to project the intrinsic
triaxial basis onto good angular-momentum states. The lowest projected bands obtained 
for the studied
$^{124}$Cs, $^{126}$Cs, $^{130}$Cs and $^{132}$Cs nuclei are shown in Figs.~1 and 2. The projected
bands displayed in the two figures are referred to as the band diagrams and are quite instructive
to unravel the intrinsic quasiparticle structures of the observed bands.
For $^{124}$Cs, it is evident from the top panel of Fig.~1 that
many configurations contribute for the yrast spectroscopy for all the studied spin values. For 
low-spin values, K=1,2 and 3  projected bands from the two-quasiparticle 
configuration with energy 2.31 MeV are almost degenerate.
For high-spin states, K=4 and 5 bands also become favourable and the strong
odd-even staggering in the K=1 band leads odd-spin members of the lowest
band resulting from the K=3 configuration. The projected bands from the other quasiparticle
state with energy equal to 2.72 MeV lie at higher excitation energies and don't become
favourable for the studied spin values.
The band diagram for $^{126}$Cs, shown in the bottom panel of Fig.~1, has lowest two bands
having K=1 and 2 in the low-spin regime, originating from the same intrinsic quasiparticle state. For 
higher-spins, it is noted that several bands contribute to the near yrast spectroscopy. Further,
due to odd-even staggering in the K=1 band, the lowest odd-spin member at high-spin has
K=3 and originate from a different quasiparticle state in comparison to the even-spin. As will
be demonstrated later, these structural changes at high-spins are reflected in the observed
transition probabilities. 

The band diagram for $^{130}$Cs, shown in the upper panel of Fig.~2, depicts K=4,5,6 and 7 
configurations, projected from the quasiparticle state with energy 0.64 MeV,
dominating the yrast spectroscopy over the whole studied spin regime. However,
it is noted that above I=19, the projected configurations from the quasiparticle state with
energy equal to 1.42 MeV are begining to play important role for the yrast states. The band
diagram for $^{132}$Cs displays quite an interesting projected band structures. For low-spin
states, the yrast band is dominated by K=4,5,6 and 7 projected from the quasiparticle
state with energy equal to 1.01 MeV, however, at
I=14, it is seen that projected K=1 band from the quasiparticle state with energy equal to
1.03 MeV crosses and becomes yrast. This crossing is quite 
interesting as the partner band in $^{132}$Cs is observed, to be discussed later, only up to I=14. 

In the third and final stage, the
projected basis, obtained above, are then employed for diagonalisation of the 
shell model Hamiltonian, Eq.~(\ref{Hamilt}). The energies obtained, after diagonalisation, for the
lowest two bands are depicted in
Fig.~3 for the four studied isotopes. The excitation energies have been subtracted by the
core contribution in order to highlight the differences. It is evident from the 
figure that the agreement
between the calculated and the measured energies is quite satisfactory. The odd-even
staggering observed for the yrast-band at lower-spin values in $^{124-126}$Cs isotopes 
is well reproduced by the 
TPSM calculations. For a better visualisation of the odd-even effect, the staggering
parameter,  $S(I)=\left[E(I)-E(I-1)\right]/2I $ is plotted in Fig.~4 for the yrast-bands
of the four studied nuclei. For the
low-spin values, the calculated staggering is larger than the observed and this discrepancy
seems to be inherent in most of the earlier theoretical investigations \cite{SY07,GR03,Jpeng03,peng03}.

It is interesting to note from Fig.~3 that in $^{132}$Cs, yrast and the first-excited
bands become almost degenerate at I=14 and the observed partner band is known only
up to this spin value. The reason for this degeneracy can be traced to the crossing
of the projected bands originating from two different quasiparticle configurations
in Fig.~1 at this spin value. To investigate the nature of the doublet-bands in $^{132}$Cs
after diagonalisation, the probability of various projected configurations in the two
wavefunctions are displayed in Fig.~5. For the yrast band, the dominant configurations
up to I=13 are K=5,6 and 7 from the quasiparticle state with energy 1.01 MeV. Above I=13,
the yrast band is composed of of the projected configurations, originating from the
quasiparticle state with energy 1.03 MeV. For the partner-band, the structure is
quite opposite to that of the yrast-band with 1.03 MeV quasiparticle state 
dominating up to I=13 and above this spin value, it is the other quasiparticle state becoming
important. Therefore, it is predicted in the present work that the yrast and the partner bands
change their character at I=14.

We shall now turn to the discussion of the electromagnetic transition probabilities, which
play a crucial role to determine whether the doublet bands originate from the breaking
of the chiral symmetry.
The transition probabilities have been evaluated using the expressions provided
in our recent work \cite{JG12} with the parameters : 
$g_l^\pi = 1, ~
g_l^\nu = 0, ~  
g_s^\pi =  5.586 \times 0.85,~ g_s^\nu = -3.826 \times 0.85 $,
and the effective charges of $e^{\pi}=1.5e$ and $e^{\nu}=0.5e$. The transition probabilities 
are depicted in Figs.~6, 7, 8 and 9 for $^{126}$Cs, $^{124}$Cs, $^{130}$Cs and $^{132}$Cs, respectively.
It is seen from Fig.~6 that both the behaviour and the magnitudes of the observed 
transitions for $^{126}$Cs 
are reproduced reasonably well by the TPSM approach. In particular, the observed 
drop in the ${B(E2)}$
transitions with spin, Fig.~6(a), is correctly reproduced. To reproduce this 
behaviour in the ${B(E2)}$ 
transitions, also observed in $^{128}$Cs, has been a drawback of the particle-rotor model 
description for these nuclei \cite{EG06}. 
The staggering of the ${B(E2)}$ transitions in the spin regime between
I=14 and 18 can be traced to the interplay between K=1 and 3 quasiparticle configurations as
is evident from the band diagram of Fig.~1. The wavefunctions after diagonalization, although,
are highly mixed, but it is clearly noted that even- and odd-spin members of the 
yrast-band have different K-decompositions.

It has been already mentioned that chiral geometry imposes stringent selection 
rules \cite{Hamamoto} on the wavefunction
with the resulting characteristics transition probabilities. For chiral geometry, 
${B(M1)}$ transitions 
are expected to depict odd-even staggering with the phase of the in-band transitions
opposite to that of inter-band transitions for the two chiral bands. This behaviour 
of the ${B(M1)}$ transitions is 
evident from Fig.~6~(c), (d), (g) and (h) in both observed as well as in the TPSM results.
Some discrepancies between the TPSM results and the
experimental values are noted in Figs.~6(d) and (g) for the side- and 
the yrast-to-side band ${B(M1)}$ transitions.

The calculated transition probabilities for $^{124}$Cs, shown in Fig.~7, depict similar
behaviour as that of $^{126}$Cs. However, as compared to $^{126}$Cs, ${B(E2)}$ depict odd-even 
staggering even for
low-spin values and the reason for this can be traced to the interplay
among several configurations in the low-spin regime as is evident from the band 
diagram plot of Fig.~1. The odd-even
staggering in the ${B(M1)}$ for this system is smaller than that of $^{126}$Cs, but
has the characteristic feature of the chiral geometry with the phase of in-band
transitions opposite to that of inter-band transitions. Transition probabilities
have not been measured for this system  and only the ratios of ${B(M1)}$/${B(E2)}$, calculated
from the intensities, are  depicted in Fig.~7.

The transition probabilities for $^{130}$Cs, displayed in Fig.~8, show B(E2) dropping with spin
for the yrast configuration and for the partner band B(E2) depict odd-even staggering since it is
composed of low-K projected states. B(M1) transitions in Fig.~8 have features as those 
expected for chiral bands. For $^{132}$Cs, electromagnetic transitions are depicted in Fig.~9 with
${B(E2)}$ for the yrast band dropping with spin as
for the other studied isotopes. The only difference is that above I=13, B(E2)'s depict
odd-even staggering and is due to dominance of the K=1 projected state above I=13. B(M1) 
transitions in Fig.~9 again have the behaviour as those expected for the chiral bands. The variation of the measured B(M1)/B(E2) ratios for the yrast band with
spin in $^{130}$Cs and $^{132}$Cs are well reproduced in the present calculations as shown in Fig. 8(e) and Fig. 9(e), respectively. Similarly, for the partner band of $^{130}$Cs, the calculated B(M1)/B(E2) ratios are in agreement with
the measured values (see Fig. 8(f)). 

\section{Summary and conclusions}
In the present work a systematic investigation of the doublet-bands observed 
in the four  Cs-isotopes with mass numbers of 124,126,130 and 132 has been 
performed using the recently developed
multi-quasiparticle TPSM approach. The transition energies of the doublet-bands are known
for all the four isotopes up to high-spin and depict similar $\gamma$-ray energies and have
been proposed to originate from the chiral symmetry breaking. However, it has
been demonstrated \cite{CM06} that the near degeneracy of the observed doublet-bands don't 
necessarily imply that the bands originate from the breaking of the chiral symmetry. The important
clues on the nature of the doublet-bands is contained in the electromagnetic transition
probabilities. It has been shown in a model study \cite{TS04} that for doublet-bands to 
originate from the breaking of the chiral symmetry, the transitions probabilities 
must obey very special selection rules. For chiral bands, it is expected that ${B(M1)}$ 
transitions depict staggering with in-band transition showing opposite phase to 
that inter-band transitions. Although, these selection rules have been established in a simplified
model, nevertheless, it is expected these should be obeyed approximately in realistic
models. 

Recently, a detailed experimental study on the electromagnetic transitions \cite{EG12} 
has been performed for the doublet-bands observed in $^{126}$Cs and this data is more exhaustive
than the previously best studied $^{128}$Cs system. It has been shown that TPSM approach 
provides a consistent
description of the energies and the measured transition 
probabilities for $^{126}$Cs. Further, it has been demonstrated that results
are consistent with the chiral interpretation for the observed doublet bands.
The predicted transition probabilities for $^{124,130,132}$Cs isotopes  also depict 
characteristic features of 
chiral symmetry breaking for the
observed bands and it is highly desirable to perform lifetime measurements for the observed 
doublet-bands in order to confirm the chiral interpretation for these systems.

\end{document}